\documentclass[conference]{IEEEtran}
\IEEEoverridecommandlockouts
\usepackage{cite}
\usepackage{amsmath,amssymb,amsfonts}

\usepackage{graphicx}
\usepackage{textcomp}
\usepackage{xcolor}
\usepackage{bm}
\usepackage{lipsum}
\usepackage[noend]{algpseudocode}
\usepackage{makecell}
\usepackage{algorithm}
\usepackage{subcaption}
\usepackage{hyperref}
\usepackage{algpseudocode}
\newcommand{\RN}[1]{%
  \textup{\expandafter{\romannumeral#1}}%
}

\makeatletter

\def\BibTeX{{\rm B\kern-.05em{\sc i\kern-.025em b}\kern-.08em
    T\kern-.1667em\lower.7ex\hbox{E}\kern-.125emX}}
\makeatletter
\newcommand*{\rom}[1]{\expandafter\@slowromancap\romannumeral #1@}

\makeatother
\begin{document}

\title{Robustness analytics to data heterogeneity in edge computing}

\author{Jia~Qian,
        Lars Kai Hansen,
        Xenofon Fafoutis,~\IEEEmembership{Member, IEEE},
        Prayag Tiwari,
        Hari Mohan Pandey

\thanks{J. Qian, L. K. Hansen and X. Fafoutis are with Department of Applied Mathematics and Computer Science, Technical University of Denmark, 2800 Lyngby, Denmark (e-mail: {jiaq, xefa, lkai\}@dtu.dk}). The research leading to these results has received funding from the European Union’s Horizon 2020 research and innovation program under the Marie Sklodowska-Curie grant agreement No. 764785, FORA-Fog Computing for Robotics and Industrial Automation. This work was further supported by the Danish Innovation Foundation through the DanishCenter for Big Data Analytics and Innovation (DABAI). P. Tiwari is with the Department of Information Engineering, University of Padova, Italy (e-mail: prayag.tiwari@dei.unipd.it). Hari Mohan Pandey is with the Department of Computer Science, Edge Hill University, Ormskirk, United Kingdom (e-mail: pandeyh@edgehill.ac.uk)}}%


\maketitle

\begin{abstract}
Federated Learning is a framework that jointly trains a model \textit{with} complete knowledge on a remotely placed centralized server, but \textit{without} the requirement of accessing the data stored in distributed machines. Some work assumes that the data generated from edge devices are identically and independently sampled from a common population distribution. However, such ideal sampling may not be realistic in many contexts. Also, models based on intrinsic agency, such as active sampling schemes, may lead to highly biased sampling. So an imminent question is how robust Federated Learning is to biased sampling? In this work\footnote{\url{https://github.com/jiaqian/robustness_of_FL}}, we experimentally investigate two such scenarios. First, we study a centralized classifier aggregated from a collection of local classifiers trained with data having categorical heterogeneity. Second, we study a classifier aggregated from a collection of local classifiers trained by data through active sampling at the edge. We present evidence in both scenarios that Federated Learning is robust to data heterogeneity when local training iterations and communication frequency are appropriately chosen.

\end{abstract}


\section{Introduction}
\begin{figure}
\centerline{\includegraphics[width=0.99\columnwidth]{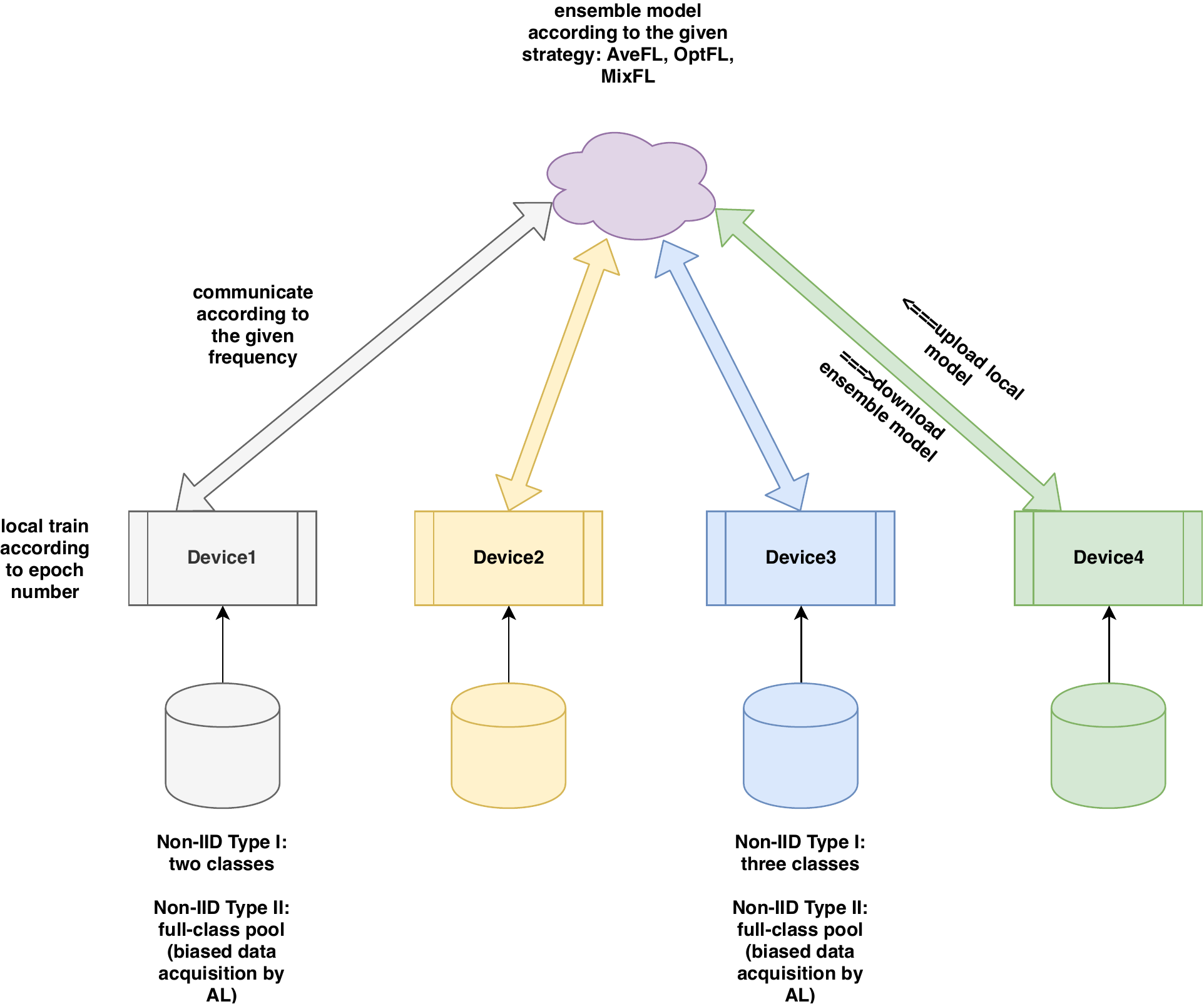}}
\caption{Federated 
Learning Scheme.}
\label{ge}
\end{figure}

Federated Learning \cite{konevcny2016federated} is a promising method to enable edge Intelligence and data protection at the same time. FL is of significant theoretical and practical interest. From a theoretical point of view, Federated Learning poses challenges in terms of, e.g., consistency (do distributed learning lead to the same result as centralized learning) and complexity (how much of the potential parallelism gain is realized). From a practical point of view, Federated Learning offers unique opportunities for data protection. In particular, Federated Learning can be realized without ``touching'' the training data, but rather the data remains in its generation location, which provides the opportunity to secure user privacy. It is very intrinsic to bring it to IoT application, in particular, when 5G is arriving. For instance, FL is exerted in industrial IoT (IIoT) to predict electric drivers' maintenance in the fog computing platform \cite{barzegaran2020fogification}. The medical data collected from distributed individuals can be processed locally and share the metadata with the central server at some point to protect personal privacy \cite{qian2020noble}. Extending FL to other machine learning paradigms, including reinforcement learning, semi-supervised and unsupervised learning, active learning, and online learning \cite{he2019central,zhao2019decentralized} all present interesting and open challenges. Some works assume that data is \textbf{Independent} and \textbf{Identically} \textbf{Distributed} (IID) on the edge devices, which is evidently a strong assumption, say in a privacy-focused application. Users are not identical; hence, we expect locally generated dataset to be the result of idiosyncratic sampling, namely, biased. We believe that data diversity is not necessarily harmful in terms of performance, which mainly attributes to the aggregation step of FL, with the condition that local training iterations and batch size are appropriately opting. A high-level depiction of this scenario is presented in Figure \ref{ge}.\\
To investigate the robustness of FL, we consider two types of Non-IID cases: Type \RN{1} we will simulate a highly biased data-generation environment, edge devices have access only to a subset of the classification classes (no overlap between them); Type \RN{2} on the edge devices, we employ AL as an active sampler to sample the most representative instances, rather than uniform sampling.\\

Our contribution can be summarized as:
\begin{itemize}
\item In general, we aim to investigate the relationship between distributed data diversity and centralized server performance in the edge computing environment.
\item More specifically, we simulate two types of biased data generation to study the robustness of FL to different unbalanced data generation level. 
\item Our experiments show that centralized server performance is highly correlated to the local training time and communication frequency. The divergent aggregation might happen if they are not appropriately chosen.
\item Finally, we investigate the effects of parameter (gradient) aggregation by comparing local neural networks activation patterns and aggregated neural networks, which shows the evidence that the server's classification capability is “inherited” from distributed devices through aggregation.

\end{itemize}
The remainder of this paper is organized as follows:
Section{~\ref{sec:related}} we will explain the preliminary concepts and introduce the related work, in Section~\ref{sec:ourapproach}, we will give the specific introduction of our scheme. In Section~\ref{sec:eval} the details of our experimental results will be recovered. Section~\ref{sec:Conc} we will conclude the paper.

For the convenience of readers, we list all the abbreviations and annotations.
\begin{table}[ht]
\caption{Abbreviations $\&$ Annotations}
\begin{center}
\begin{tabular}{|c|c|}
\hline
Abbreviations & Full Name\\
\hline
FL & Federated learning\\
\hline
Al & Active Learning\\
\hline
IID & Independent Identical Distributed\\
\hline
AveFL & Average Federated Learning\\
\hline
OptFL & Optimal Federated Learning\\
\hline
MixFL & Mixed Federated Learning\\
\hline
Mc-drop & Monte carlo dropout\\
\hline
Non-IID Type \RN{1} & Active Learning sampler\\
\hline
Non-IID Type \RN{2} & Non overlap between categories\\
\hline

\end{tabular}
\label{abbreviation}
\end{center}
\end{table}

\section{Preliminaries and Related Work}
\label{sec:related}

\subsection{Federated Learning}

FL uses a coordinated fashion to train a global model by dynamically collecting models from distributed devices for some rounds. It was first proposed by \cite{konevcny2016federated} for the user privacy consideration in mobile networks, and it is a very practical framework in edge computing. \cite{diro2018distributed} employs FL to detect attacks in a distributed system, \cite{lakshminarayanan2017simple} predicts model uncertainty by a deep aggregated model, and \cite{zhu2019multi} aims to optimize the structure of neural network in FL. Some FL-based applications assume the data is IID on edge devices. \cite{zhao2018federated} considers Non-IID data, but it focuses on the observation that accuracy reduction caused by Non-IID is correlated to weight diversity.
Our work extends it, studying two types of Non-IID data: (i) Type \RN{1} we will simulate a highly biased data-generation environment, whereby edge devices can only generate their categories without any overlap between devices, (ii) Type \RN{2} whereby we employ AL on the edge devices as an active sampler to simulate a slightly biased data generation. 


\subsection{Active learning of neural networks}
Labeling is challenging and expensive when data generation increases exponentially. Thus, when intelligence sits close to edge users, it is therefore natural to utilize the interaction between machines and users/humans. We combine Federated Learning (FL) and Active Learning (AL) as Non-IID Type \RN{2}, and we reported the prototype in \cite{qian2019active}. Theoretically, AL may achieve one of the following situations: higher accuracy with the same amount of data or with a given performance using fewer data. According to the formula of incoming data, it can be grouped as pool-based and stream-based. The stream-based AL approach is used when the data arrives in a stream way, and the model must decide whether to query from the ``Oracle'' or discard it.

\begin{figure}
\centerline{\includegraphics[width=0.9\columnwidth]{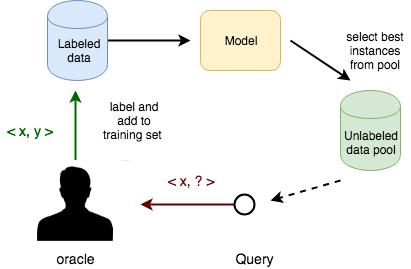}}
\caption{Pool-based Active Learning Scheme.}
\label{pool_framework}
\end{figure}

The pool-based approach (Figure \ref{pool_framework}) is composed of an initially trained model, an ``Oracle'', an unlabeled data pool, and a small labeled dataset. More specifically, the initially trained model elaborately opts for some representative samples out from the unlabeled pool based on the acquisition function. After that, it asks the oracle to label them and includes the labeled ones to the training set for future training. We can repeat such operations for several times. In the previous work \cite{qian2019active,qian2019distributed}, we train our model whenever lately-labeled data is added, along with the old (labeled) data. In this paper, we consider Online Active Learning, which means we immediately discard the data after training (more details in Section \ref{noniid2}), but with a negligible small subset shared across the devices. To apply AL on a neural network, we firstly build a Bayesian Neural Network (BNN), which can be considered as a model that outputs different values for the same input fed in the model several times. There is no analytical form of the posterior distribution in the neural network; typically a surrogate distribution q is introduced to approximate it by minimizing the distance between them. We implement it through a free ride Dropout \cite{srivastava2014dropout}, feeding the same input multiple times to approximate a distribution with certain mean and variance. It can be proved that running dropout is approximated to apply Bernoulli prior on the model parameters. More details can be found in \cite{qian2019active}.
\subsection{Boosting Approach}
Boosting is designed to improve any machine learning method, e.g., tree-like classifiers, by aggregating many weak learners through bias and variance reduction \cite{breiman1996bias} \cite{zhou2012ensemble}. The approach of the present work can be related to boosting by viewing the aggregation as a combination of 'weak' (specialized) edge models in repeated steps of the federation. In \cite{chen2015xgboost}, the authors proposed the Boosting Gradient Classifier, which has a set of weak learners and sets off by creating a weak learner, and it keeps increasing after every iteration. The set of learners is built by randomly combining features. It seeks an appropriate combination $\hat{F}$ of $f_i$ such that approximates the true $F$, expressed as $\hat{F}(x) = \sum_{i=1} \beta_i f_i(x)$. Apart from computing gradients during training, it also computes the second-order derivative to decide the learning rate. Instead, our method keeps the number of weak learners constant, which is the number of edge devices. Analogously, we can also make it dynamic, like boosting gradient classifiers. Another difference is that we do not compute the second-order derivatives to decide the learning rate; instead, we empirically choose one as the Neural Network has a large amount of parameters. The Boosting Gradient classifier typically has a good performance in conventional machine learning application \cite{xu2014gradient,klein2011boosting,son2015tracking}. The main steps of Boosting Gradient are as follows:
\begin{itemize}
\item if $m=0$, we output the prediction by average the outcomes from the weak learners $\hat{F}(x)=\frac{1}{n}\sum_{i=1}^n y_i$.
  \item For iteration m from 1 to M (the case $m \neq 0$):
    \begin{itemize}
     \item model in iteration m defines as
      \[W_{m} = W_{m-1} + \gamma_{m} g_m(x)\]
     \item $\gamma_{m}$ and $g_m(.)$ are computed separately by the first and second order of loss function L.
      \[g_m(x)=-[\frac{\partial L(y_{i},\hat{F}(x,W_m))}{\partial W_m}]_{W=W_{m-1}} \]
      \[ \gamma_{m}= [\frac{\partial^2 L(y_{i},\hat{F}(x_{i},W_m))}{\partial W_m^2}]_{W=W_{m-1}} \]
     \end{itemize}
  $F_{m-1}$: the collection of learners up to stage m-1.\\
  $\gamma_{m}$: the learning rate in iteration m.\\
  $g_{m}(x)$: gradient in stage m. \\
\end{itemize}
 In summary, both Boosting Gradient classifier and FL attempt to improve the performance by assembling a set of weak models. The Boosting Gradient classifier works on a dataset with extracted features, specifically, optimize the learning rate and keeps the number of weak learners increasing; whereas we design federated learning for neural network, the learning rate is empirically decided due to the computation problem and the size of models is constant, we can make it dynamic though.
 \subsection{Other works}
 Data non-IID was introduced in \cite{zhao2018federated}, and they tackle it by introducing a relatively small global subset that may somehow capture the whole distribution, shared across all devices. Similarly, \cite{nishio2019client} suggests using data distillation to extract a low-dimension (or sparse) representation of the original data. However, it is computationally expensive; in particular, it is typically carried out at the edge side where only little computation resources can be offered. \cite{kairouz2019advances} converts non-IID data distribution as an advantage by considering it as a multi-task optimization, which conforms to our conclusion. Furthermore, \cite{hashimoto2018fairness} utilizes distributionally robust optimization to minimize the worst-case risk over all the distributions close to the empirical distribution. 


\section{Proposed scheme}\label{sec:ourapproach}

\subsection{Federated Learning Aggregation Strategies}
More specifically, let's assume the server shares the model (at round t) $W_t$ with n devices for their local updating, and the updated models are denoted as $W_t^1, W_t^2, W_t^3,...W_t^n$. Then, the devices upload the improved models to the server, and the server outputs the aggregated model according to the following criterion:
\begin{equation}
W_{t+1} := \sum_i^n \alpha_{i} * W_t^i \,
\label{eq1}
\end{equation}

The combination weights $\alpha_i$s can be uniformly distributed or determined to reflect network performance. The former is referred to  as \textit{AveFL} (Algorithm~\ref{alg:avefl}). The learning process is iterative. We also consider the second scheme, where we opt for the highest-accuracy model, namely, set $\alpha^*$ of the best model equal to one, and the rest to zero, labeled as \textit{OptFL} (Algorithm~\ref{alg:optfl}). In Section \ref{sec:eval}, we evaluate the schemes and a combination of \textit{AveFL} and \textit{OptFL}, named as \textit{MixFL}. The latter selects the best model of the former two (Algorithm~\ref{alg:mixfl}).


\begin{algorithm}[t]
\caption{\textit{AveFL}}
\label{alg:avefl}
\begin{algorithmic}[1]
\State \text{Input: $W_j^t$: local models at round $t$ }
\State \text{Output: ensemble model $W^t$}
\State \text{$W^t =\frac{1}{n} \sum_{j=1}^n W_j^t$}
\end{algorithmic}
\end{algorithm}

\begin{algorithm}[t]
\caption{\textit{OptFL}}
\label{alg:optfl}
\begin{algorithmic}[1]
\State \text{Input: $W_j^t$: local models at round $t$, $A(.)$: measure accuracy}
\State \text{X: test dataset}
\State \text{Output: ensemble model $W^t$}
\State \text{$W^t = \text{argmax} A(\text{BNN}(X, W_j^t))$ for $j=1,2,..,n$ }
\end{algorithmic}
\end{algorithm}

\begin{algorithm}[t]
\caption{\textit{MixFL}}
\label{alg:mixfl}
\begin{algorithmic}[1]
\State \text{Input: local models at round $t$ $W_j^t$}
\State \text{Output: ensemble model $W^t$}
\State \text{$W_{\text{ave}}^t = \textit{AveFL}(W_j^t)$ }
\State \text{$W_{\text{opt}}^t = \textit{OptFL}(W_j^t)$}
\State \text{$\text{acc}_{\text{ave}}=A(\text{BNN}(X,W_{\text{ave}}^t))$}
\State \text{$\text{acc}_{\text{opt}}=A(\text{BNN}(X,W_{\text{opt}}^t))$}
\If{$\text{acc}_{ave} >= \text{acc}_{opt}$}
\State \text{Return $W_{\text{ave}}^t$}
\Else
\State \text{Return $W_{\text{opt}}^t$}
\EndIf
\end{algorithmic}
\end{algorithm}


\begin{figure}
\centerline{\includegraphics[width=0.8\columnwidth,trim={2.5cm 1cm 2.3cm 1cm},clip]{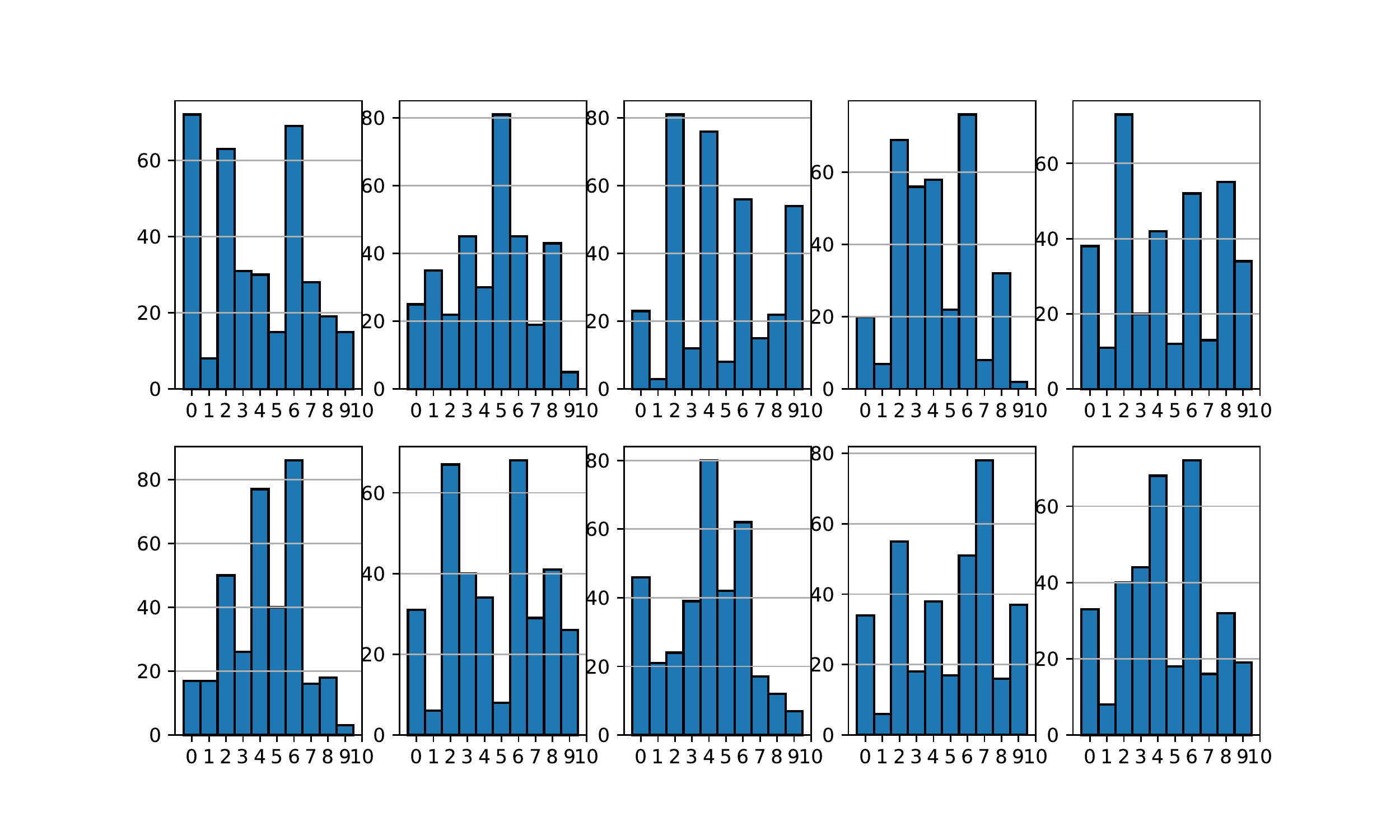}}
\caption{Biased Data Acquisition by Active Learning: we demonstrate the distribution of data acquired by AL for 10 acquisitions. They are unbalanced in different ways for every acquisition.}
\label{al_hist}
\end{figure}

Rather than ensemble the weights of models in Equation~\ref{eq1}, we can also work on the gradients. We conclude that one-batch weight average is equal to gradient average. Suppose we have $n$ devices, and training data $D$ ($|D|=N$) is sectioned into $n$ parts as $D_1,D_2,...D_n$, $|D_1|=N_1,|D_2|=N_2,...|D_n|=N_n$. The corresponding weights inferred from $D_i$ is $W_i$. Then we define a cost function $G(D)=\sum_{i=1}^N g(\widetilde{y}_i,y_i,w)$ and the initial model is $W_0$, $\beta$ is the learning rate. We first define average one-batch weights of models as shown in Equation~\ref{non-uniform2}, notably, the local update of edge devices is after one batch (no iteration of the batch), otherwise it is \textbf{not} $W_0$ in cost function $g(.)$. The gradient ensemble is defined in Equation~\ref{non-uniform3}.

\begin{equation}
        \sum_{i=1}^n \alpha_i = 1
 \label{non-uniform1}
\end{equation}

\textbf{Aggregated Weights:}\\
\begin{equation}
    \begin{split}
        W := \sum_{i=1}^n \alpha_i (W_0+\beta G(D_i))\\
        = \sum_{i=1}^n \alpha_i(\underbrace{W_0+\beta \frac{1}{N_i}\sum_{j=1}^{N_i} g(\widetilde{y}_j,y_j,\mathbf{W_0})}_\text{updated W from device i})\\
        = W_0 \sum_{i=1}^n \alpha_i + \beta  \sum_{i=1}^n \alpha_i \frac{1}{N_i} \sum_{j=1}^{N_i} g(\widetilde{y}_j,y_j,W_0)\\
        = W_0 + \beta  \sum_{i=1}^n \alpha_i \frac{1}{N_i} \sum_{j=1}^{N_i} g(\widetilde{y}_j,y_j,W_0)
    \end{split}\label{non-uniform2}
\end{equation}

\textbf{Aggregated Gradients:}\\
\begin{equation}
    \begin{split}
        W := W_0 + \beta * (\sum_{i=1}^n \alpha_i G(D_i))\\
        = W_0 + \beta (\sum_{i=1}^n \alpha_i \underbrace{\frac{1}{N_i}\sum_{j=1}^{N_i} g(\widetilde{y}_j,y_j,W_0)}_\text{gradient of device i})
    \end{split}\label{non-uniform3}
\end{equation}


In each iteration, keeping $W_0$ the same for all edge devices is a mandatory step; otherwise, weight divergence can occur. If initial models are different, they might be placed in a different low-cost region of the cost landscape. Thus, after the average aggregation step, it might be sub-optimal. In this paper, we also aim to investigate how the number of local training influences the result, and we decide to work on the weights aggregation for the sake of convenience.

\subsection{Method for Non-IID Type \rom{1}}\label{sec:ourapproach2}
Our approach can be divided into two stages: local learning and ensemble. The two stages will be iterated in one round. 

\begin{enumerate}
\item \textbf{Initialization:} At initialization, the centralized server trains an initial model $W^0$  using m data samples. We denote the model as $W^t$, where $t$ is the number of current round.
\item \textbf{Distribution:} Server dispatches the model $W^t$ to $n$ activated edge devices $D_1,D_2,...,D_n$.
\item \textbf{Local Training:} All edge devices implement local training and update their models $W^t_1, W^t_2,..., W^t_n$. This step incorporates one or multiple cycles of data acquisition.
\item \textbf{Ensemble:} Edge devices transmit their corresponding models to server and the server ensembles $W^t_i, i=1,2..,n$ to get $W^{t+1}$. The ensemble could be \textit{AveFL}, \textit{OptFL} or \textit{MixFL}.
\item Repeat steps 2-4 if necessary.
\end{enumerate}
The specific algorithm is described in Algorithm~\ref{alg:typeI}.

\begin{algorithm}[t]
\caption{Non-IID Type \rom{1}}
\label{alg:typeI}
\begin{algorithmic}[1]
\If {t==0}
\State \text{set initial training images number} $m=400$ 
\State \text{form initial training set} $X_{0}$ with size equal to $m$
\State \text{train model} $W_{0}$ by $X_{0}$
\State \text{dispatch model} $W_{0}$ to device $D_1,D_2,D_3,...,D_n$
\Else 
\For{j=1,2,..,n} \text{n devices (in Parallel)}
\State $X_{train}^t=\text{RandomSample()}$ 
\State $\text{W}_{j}^t=\text{Train}(X_{train}^t)$
\EndFor
\State \textbf{end}
\State \text{$W^{t+1}$ = \textit{AveFL}($\text{W}_{j}^t$) for j=1,2,...,n}
\State $\text{W}_{j}^{t+1}=W^{t+1}$
\State \textbf{end}
\EndIf
\State \textbf{end}
\end{algorithmic}
\end{algorithm}

\subsection{Method for Non-IID Type~\rom{2}}\label{noniid2}

\begin{algorithm}[h]
\caption{Non-IID Type~\RN{2}}
\label{non-iid2}
\begin{algorithmic}[1]
\State \text{\textbf{Input:} $X_1\cup X_2\cup X_3,..\cup X_n$, $X_{\text{ini}}, W^{0}, k$}
\State \text{\textbf{Return:} model  $f(W)$}
\If{t==0}
\State \text{$W^1 \leftarrow W^0 - \alpha \nabla f(X_{\text{ini}};W^0)$}
\Else
\For{t=1,2,..T}
\State \textbf{$=>$Devices:}
\For{j=1,2,...,n} \text{n devices (in Parallel)}
\State \text{$\log p_j,p_j = \text{BNN}(f_j(W^{t}),x_j)$}
\State \text{compute entropy: $S_j = -p_j \times \log p_j$}
\State \text{sort in descending order and pick top k: }
\State \text{$D^t_j = \text{sort}(S_j)[k]$}
\State \text{$W^{t+1}_j = W^t_j-\alpha \times \nabla f(D^t_j;W^t_j)$}
\EndFor \State \textbf{end}
\State \textbf{$=>$Server:}
\State \text{Aggregation: $W^{t+1}$ = \textit{AveFL}($\text{W}_{j}^{t+1}$) for j=1,2,...,n}
\EndFor \State \textbf{end}
\EndIf \State \textbf{end}
\end{algorithmic}
\end{algorithm}

\begin{algorithm}
\caption{\textit{Bayesian Neural network} (BNN)}
\label{BNN}
\begin{algorithmic}[1]
\State \text{\textbf{Input:} $f_i(W^t),x_i$}
\State \text{\textbf{Return:} $\log \overline{p},p$}
\State \text{$s=0$}
\For{g = 1,2,..r}
\State \text{$p=f_i(x_i;W^t)$}
\State \text{$s += p$}
\EndFor
\State \textbf{end}
\State \text{$\overline{p} =\frac{1}{r}\times s$}
\end{algorithmic}
\end{algorithm}


We consider FL with AL as Non-IID Type \RN{2} since AL samples a subset of data with higher uncertainty, which leads to biased samples. First, we divide the whole training set into four parts, one part for one edge device. Then we build a pool with 4000 images randomly sampled from one part for the computation consideration since we need to measure the uncertainty of every data point in the pool. The pool is almost balanced; however, the batches generated by the active sampler is unbalanced, one example of 10 acquisitions shown in Figure \ref{al_hist}. For scalability, in this paper, we perform an online AL. Namely, the model is further trained only by the new batch, without the access of the old data (except small subset with 50 images), which is different from the previous work that we train all the data from scratch whenever new data is coming. We try to alleviate the forgetting problem of online learning by a cheap trick, storing 50 images, a balanced set (5 images per class) and will be combined with a new batch to train the model. After completing current-round training, we dump the new batch and only keep 50 images in the labeled set. Moreover, we also use weight decay \cite{krogh1992simple} as a regularizer that prevents the model from changing too much. We define it in Equation \ref{weight_decay}, $E(.)$ is the cost function, $w^t$ is the model parameter at round $t$ and $\lambda$ is a parameter governing how strongly large weights are penalized.

\begin{equation}\label{weight_decay}
    E(w^{t+1})=E(w^t) + \frac{1}{2}\lambda\sum_i (w_i^t)^2
\end{equation}

\cite{pinsler2019bayesian} learns the weights that can mostly approximate the distribution of the data from the pool by solving an optimization problem. It is highly computation-demanding and not suitable for edge devices. Another work \cite{sarwar2019incremental} attempts to avoid forgetting by dividing the NN architecture into parts and assigning them to different edge devices, but it requires restricting synchronization during aggregation.
Our Non-IID Type~\RN{2} method is sketched as:
\begin{enumerate}
\item \textbf{Initialization:} In the beginning, a centralized server trains an initial model $W^0$  using m data samples. Without the loss of generality, we denote the model by $W^t$, where $t$ is the current round.
\item \textbf{Sharing:} The central server shares the model $W^t$ to $n$ activated edge devices $d_1, d_2,..., d_n$.
\item \textbf{Local Training:} All edge devices implement AL on a Bayesian Neural Network approximated by Dropout \cite{srivastava2014dropout}, locally train and update their models $W^t_1, W^t_2,..., W^t_n$. This step incorporates one or multiple cycles of data acquisition.
\item \textbf{Aggregation:} Edge devices transmit their corresponding models to server and the server aggregates $W^t_i, i=1,2..,n$ to get $W^{t+1}$. The aggregation step could entail the average, performance-based or mixed mechanisms.
\item Repeat steps 2-4 if necessary.
\end{enumerate}
The specific algorithm is described in Algorithm~\ref{non-iid2}.

\subsection{Architecture}

Our model consists of four convolutional layers, one fully-connected layer and a softmax layer shown in Table~\ref{tab2}. 
\begin{table}[ht]
\caption{Neural Network Architecture}
\begin{center}
\begin{tabular}{|c|c|c|c|}
\hline
\textbf{layer}&\textbf{layer name} & \makecell{output channels or \\number of nodes} & \textbf{kernel size}\\
\hline
\textbf{\makecell{1}} & \textbf{Conv2d} & 64 &4x4\\
\hline
\textbf{\makecell{2}} & \textbf{ReLu} & -&-\\
\hline
\textbf{\makecell{3}}& \textbf{Conv2d}  & 16&5x5\\
\hline
\textbf{\makecell{4}}&\textbf{ReLu} & - &- \\
\hline
\textbf{\makecell{5}} &\textbf{Max Pooling}& -&2x2\\
\hline
\textbf{\makecell{6}} &\textbf{Dropout}& - &0.25\\
\hline
\textbf{\makecell{7}}&\textbf{conv2d}&32&4x4\\
\hline
\textbf{\makecell{8}} & \textbf{ReLu} &-&-\\
\hline
\textbf{\makecell{9}} &\textbf{conv2d}& 16 &4x4\\
\hline
\textbf{\makecell{10}}&\textbf{ReLu} & - &- \\
\hline
\textbf{\makecell{11}} &\textbf{Max Pooling}& -&2x2\\
\hline
\textbf{\makecell{12}} &\textbf{Dropout}& - &0.25\\
\hline
\textbf{\makecell{13}} &\textbf{Linear}& 128 &-\\
\hline
\textbf{\makecell{14}}&\textbf{ReLu} & - &- \\
\hline
\textbf{\makecell{15}} &\textbf{Dropout}& - &0.5\\
\hline
\textbf{\makecell{16}} &\textbf{Output}& 10 &-\\
\hline
\end{tabular}
\label{tab2}
\end{center}
\end{table}
Note that we did not use batch normalization \cite{ioffe2015batch} in the architecture since the biased batch normalization has a deleterious effect on the aggregation performance. Mathematically, it defines as shown in Equation~\ref{bn1} and Equation~\ref{bn2}. Suppose we have a batch $B=\{x_j\}_{j=1,2..,m}$, then it is normalized by its mean $\mu_B$ and variance $\sigma_B$ (computed in Equation (\ref{bn1})), and then we infer a new mean ($\beta$) and variance ($\gamma$) during training process. It may reduce the internal covariate shift and speed up the training procedure to form new representation of data (Equation (\ref{bn2})). \\
In the Non-IID case, the means ($\beta$) and variances ($\gamma$) optimized in the local training stage are decided by their biased data, and it is not beneficial during the aggregation stage in our experiments. It is very critical to enable aggregation effect in highly biased data generation; otherwise, the aggregated model performs very poorly (e.g., $20\%$ accuracy with batch normalization and $47\%$ otherwise).
\begin{equation}\label{bn1}
\mu_B = \frac{1}{m}\sum_{i=1}^m x_i, 
\sigma_B^2 = \frac{1}{m}\sum_{i=1}^m (x_i-\mu_B)^2
\end{equation}

\begin{equation}\label{bn2}
 \hat{x}_i = \frac{x_i-\mu_B}{\sqrt{\sigma_B^2}+\eta},
y_i = \gamma \hat{x}_i + \beta   
\end{equation}


\section{Experiments and Results}\label{sec:eval}
In this section, we discuss the experimental setup along with the dataset used for our evaluation, and the results of these experiments.

\subsection{Real Dataset}

\begin{figure}
\centerline{\includegraphics[width=90mm]{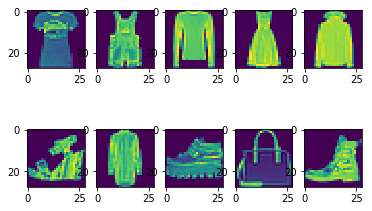}}
\caption{Fashion MNIST dataset: 10 classes, every image has $28\times28$ pixels. }
\label{fmnist}
\end{figure}

Fashion-MNIST (shown in Figure \ref{fmnist}) is one benchmark image dataset published by Zalando, as the alternative of MNIST dataset. It is formed by a training set with 60,000 examples and a test set with 10,000 examples and 10 classes. One image has $28\times28$ pixels for width and height and one channel. Each pixel value ranges between 0 and 255, indicating the shades of grey. 

\subsection{Non-IID Type \RN{1}}
We first evaluate the case of Non-IID Type \RN{1}: we have a ten-classes dataset and four edge devices (D1, D2, D3, and D4), randomly assign two classes to two devices and three classes to another two devices without overlap. More specifically, class 0 and 1 were assigned to D1, class 2 and 3 to D2, class 4, 5, and 6 to D3, and 7, 8, 9 to D4. Note, if we train a single neural network sequentially: first on the subset of classes 0 and 1, then on classes $2,3$, next  $4,5,6$, and finally $7,8,9$, the model would suffer catastrophic forgetting. It will forget most of the patterns learned before, and capable of classifying the class corresponding to the last subset (around $28\%$).  

\subsubsection{Epochs}
One of the most critical hyper-parameters is the amount of local training before aggregation on the centralized server. In this work, we redefine the concept of `epoch' since it usually refers to the number of times the learning algorithm will work through the whole training dataset. Here we consider mini-batch gradient descent; thus, `epoch' refers to the number of times the algorithm goes through the mini-batch. 
\begin{figure}[t!] 
\begin{subfigure}{0.249\textwidth}
\includegraphics[width=\linewidth]{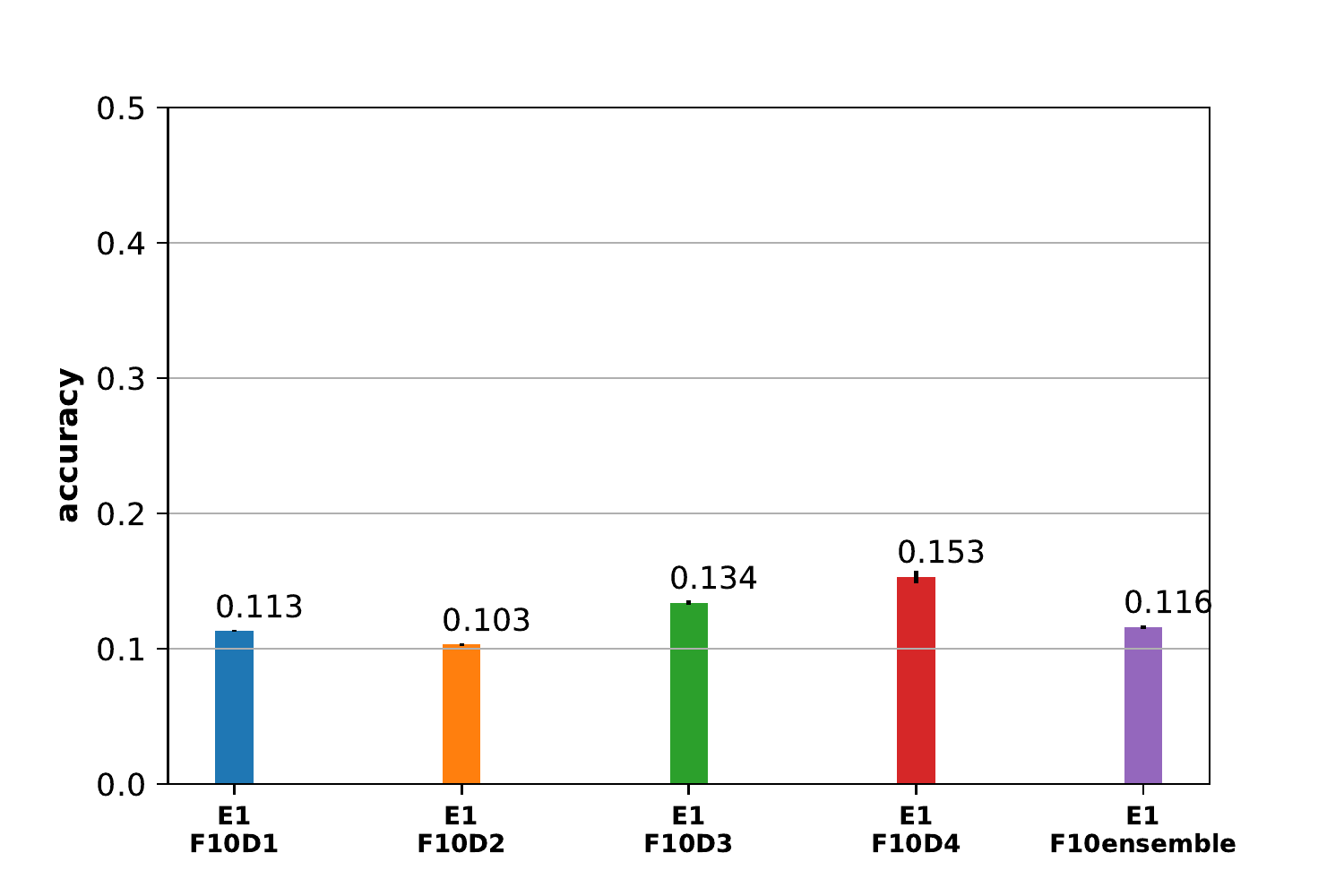}
\caption{Epoch: 1} \label{epoch:a}
\end{subfigure}\hspace*{\fill}
\begin{subfigure}{0.249\textwidth}
\includegraphics[width=\linewidth]{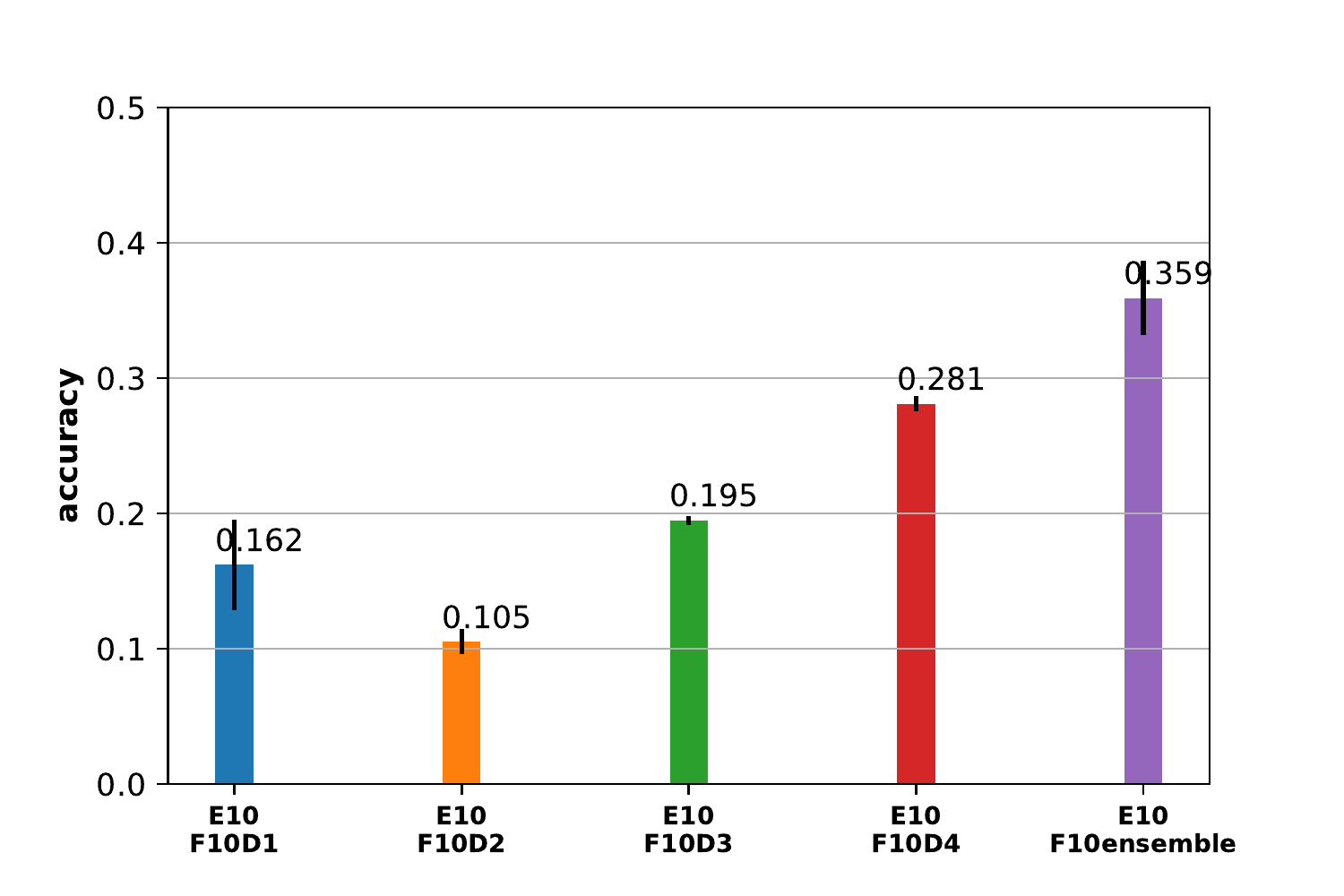}
\caption{Epoch: 10} \label{epoch:b}
\end{subfigure}
\medskip
\begin{subfigure}{0.249\textwidth}
\includegraphics[width=\linewidth]{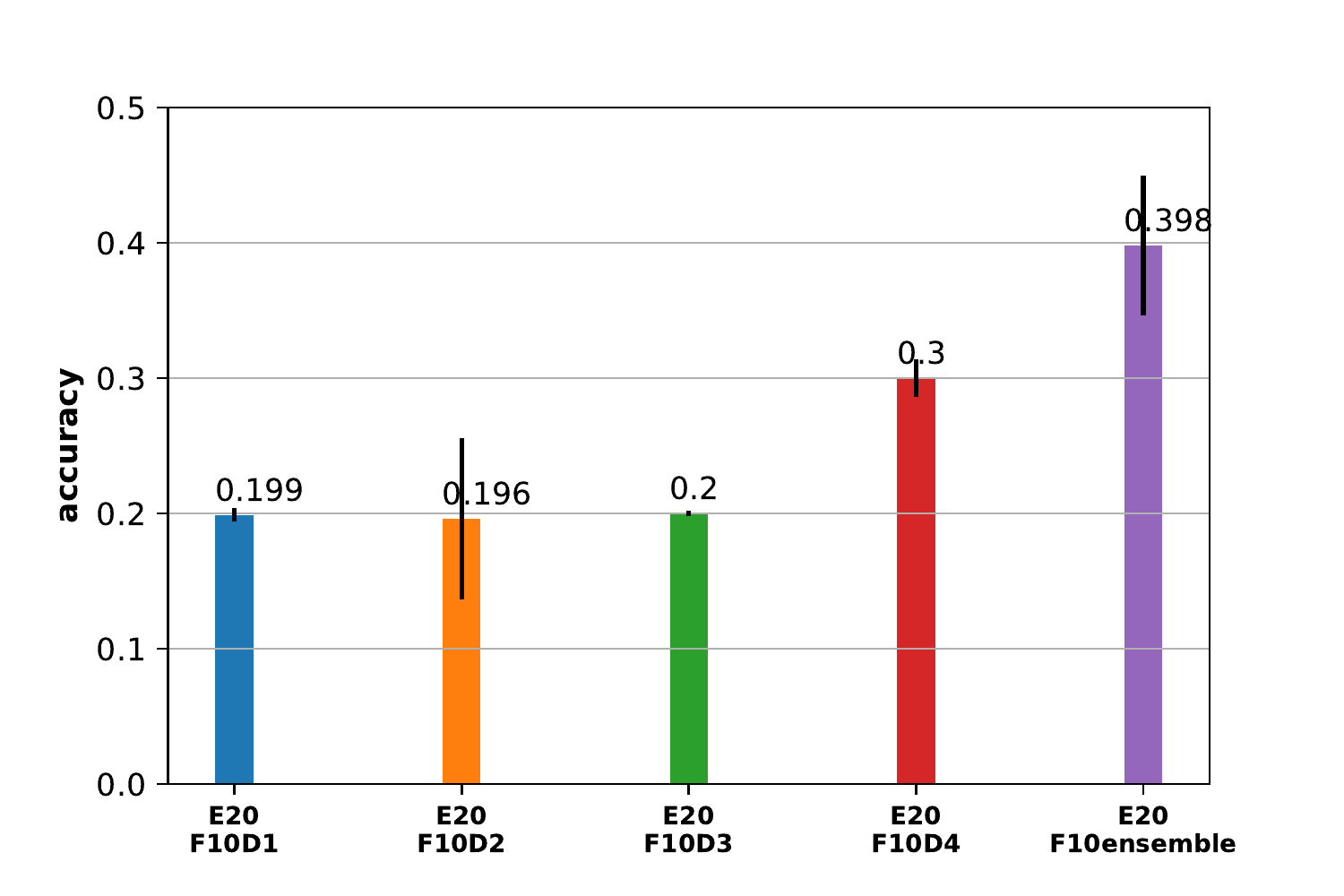}
\caption{Epoch: 20} \label{epoch:c}
\end{subfigure}\hspace*{\fill}
\begin{subfigure}{0.249\textwidth}
\includegraphics[width=\linewidth]{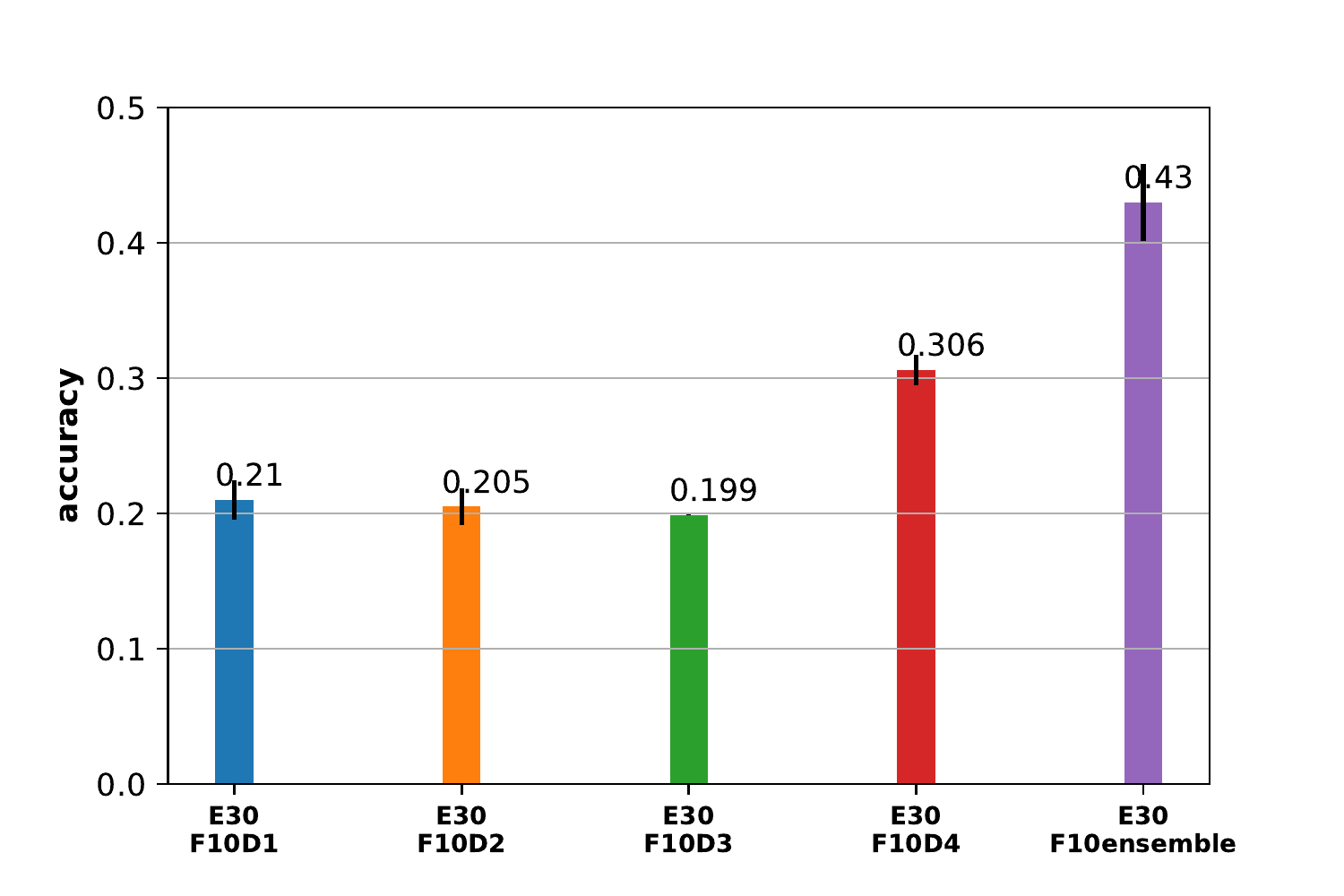}
\caption{Epoch:30} \label{epoch:d}
\end{subfigure}
\medskip
\begin{subfigure}{0.249\textwidth}
\includegraphics[width=\linewidth]{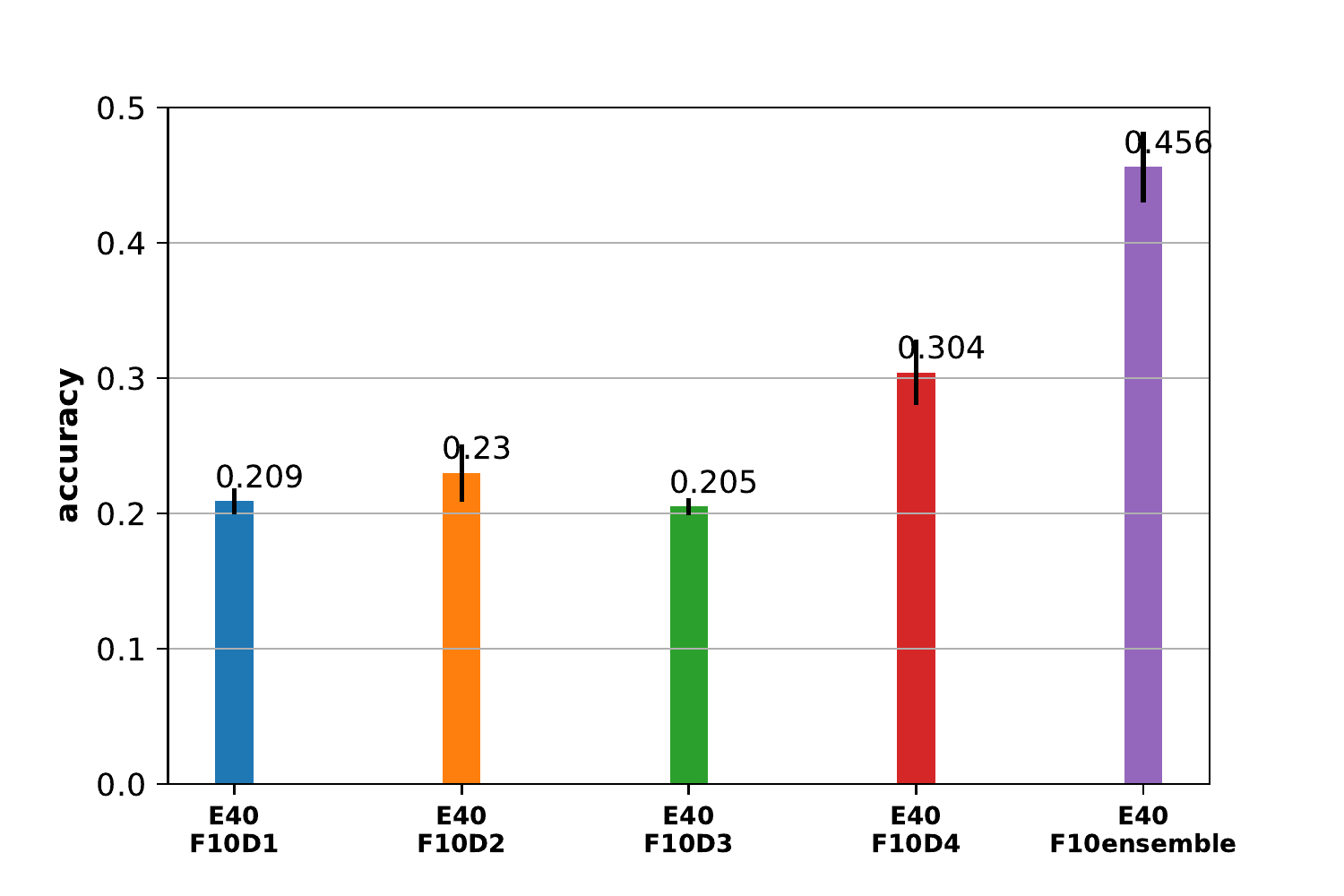}
\caption{Epoch: 40} \label{epoch:e}
\end{subfigure}\hspace*{\fill}
\begin{subfigure}{0.249\textwidth}
\includegraphics[width=\linewidth]{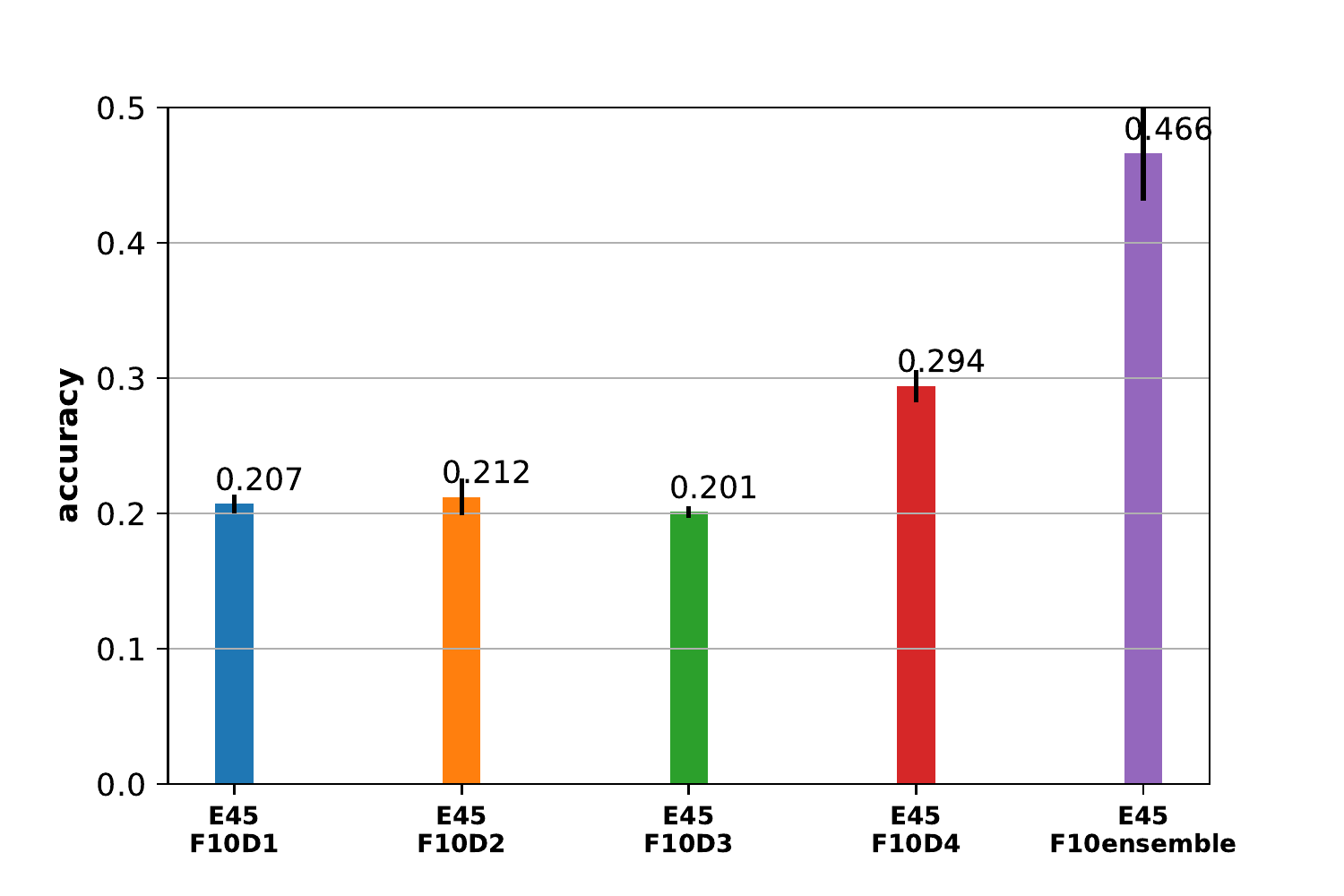}
\caption{Epoch: 45} \label{epoch:f}
\end{subfigure}
\caption{\textbf{Epoch Analysis:} The various cases are labeled using the format `ExFyDz', where
`E' is the epoch, `F' is the ensemble frequency, and `D' is the device identifier or the ensemble model respectively. For a given batch, the number of epochs during local training highly influences the ensemble performance. The experimental results show that we should ensure  sufficient difference between local models to enable the ensemble effect, which is also related to divergence study in the following experiment.}
\label{ep_com}
\end{figure}

\begin{figure*}[t]
\centerline{\includegraphics[width=0.8\linewidth]{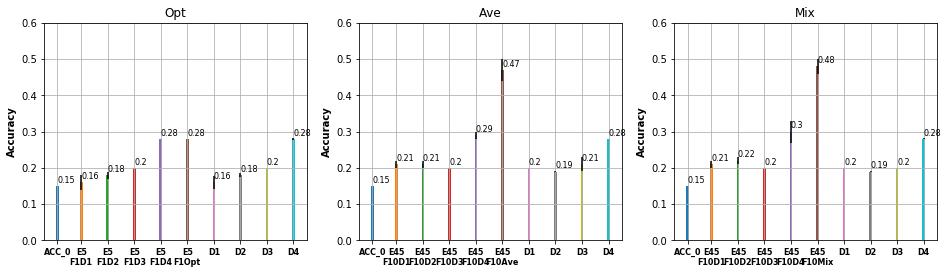}}
\centerline{\includegraphics[width=0.8\linewidth]{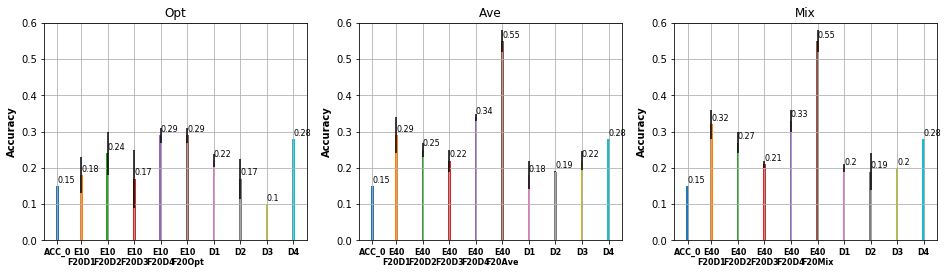}
}
\caption{\textbf{Ensemble strategies:} We compare accuracy with different ensemble strategies, namely \textit{AveFL}, \textit{OptFL} and \textit{MixFL} (top: 10 acquisitions, bottom: 20 acquisitions).
The various cases are labeled using the format `ExFyDz', where
`E' is the epoch, `F' is the ensemble frequency, and `D' is the device or the ensemble model respectively.
ACC\_0 is the initial accuracy. 
The rightmost four bars are the performance by the independent local models without considering ensemble.}
\label{fig4}
\end{figure*}

As shown in the Algorithm \ref{alg:typeI}, at the beginning of every round, all the devices have the identical model $W^t$, the number of epochs will decide how much variance between updated models $W_1^t, W_2^t,.., W_i^t$, which is produced by one batch (or multiple batches, decide by ensemble frequency that we will discuss later). If the epoch number is not big enough, after ensemble there will not have a big difference. In Figure~\ref{ep_com}, we plot the accuracy of four distributed models and the ensemble model. Among them, the leftmost four bars represent the accuracy of local models  and the rightmost one is the accuracy of the ensemble model. Note the initial accuracy is $15\%$. Figure~\ref{epoch:a} to Figure \ref{epoch:f} correspond to different epoch numbers, the accuracy almost monotonically increase with the increment of epoch number, no only for ensemble performance but also for local models. In figure \ref{epoch:f}, after enough training, three local models reach the highest accuracy they can, $20\%,20\%,30\%$ as they own two, two and three classes correspondingly.

\begin{figure}[t!] 
\begin{subfigure}{0.249\textwidth}
\includegraphics[width=\linewidth]{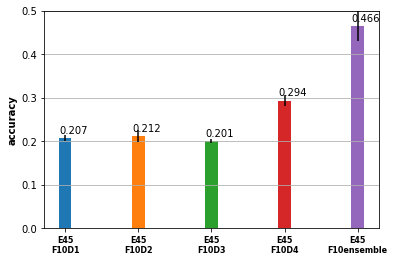}
\caption{Frequency: 10} \label{freq:a}
\end{subfigure}\hspace*{\fill}
\begin{subfigure}{0.249\textwidth}
\includegraphics[width=\linewidth]{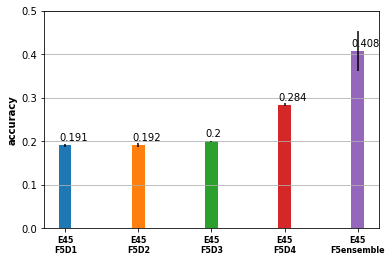}
\caption{Frequency: 5} \label{freq:b}
\end{subfigure}
\medskip
\begin{subfigure}{0.249\textwidth}
\includegraphics[width=\linewidth]{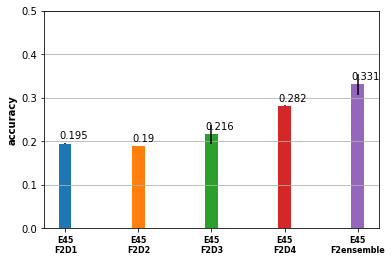}
\caption{Frequency: 2} \label{freq:c}
\end{subfigure}\hspace*{\fill}
\begin{subfigure}{0.249\textwidth}
\includegraphics[width=\linewidth]{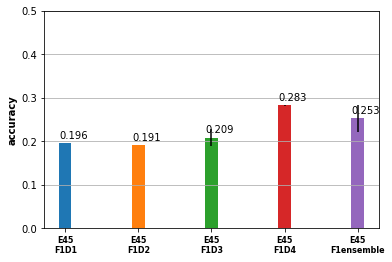}
\caption{Frequency: 1} \label{freq:d}
\end{subfigure}
\caption{\textbf{Ensemble Frequency Analysis:} 
The various cases are labeled using the format `ExFyDz', where
`E' is the epoch, `F' is the ensemble frequency, and `D' is the device or the ensemble model respectively.
For a given Epoch 45 and 10 acquisitions of data, we plot the performance with respect to different ensemble frequencies
High ensemble frequency has higher accuracy, increasing the cost of communication, and vice versa.}
\label{noniid-f}
\end{figure}

\subsubsection{Aggregation Frequency}
In \cite{qian2019active}, we only consider one-shot FL (aggregate only once), here we also study the aggregation frequency, which defines the number of acquisitions to train during local training. For instance, assume that we have $10$ acquisitions (fixed budget, 400 data for every acquisition), if aggregation frequency is $5$, it means that every $10/5=2$ acquisitions we aggregate. Note that aggregation frequency is different from epoch: epoch defines the number of repetitions given the acquisition number (training data size), whereas aggregation frequency decides the number of acquisitions, though, both of them are critical factors to enable the performance. If the aggregation frequency is low, we aggregate after a relatively large number of training data, it reduces the communication cost and takes the risk of severe divergence. Instead, if the aggregation frequency is high, we aggregate after a small amount of data, we can avoid the divergence problem, but with increasing the cost of communication. For a given epoch number 45, in Figure~\ref{noniid-f} we demonstrate the results corresponding to different aggregation frequencies. Correspondingly, we plot the performance concerning different aggregation frequencies (10, 5, 2, and 1). 
From Figure~\ref{freq:a} to Figure~\ref{freq:d}, the aggregated accuracy decreases with the decrements of aggregation frequency. We will look at this problem from analyzing the weight divergence in Section~\ref{weight_divergence}. Both the epoch and the aggregation frequency cause divergence, but aggregation has more significant impact than the epoch number.

\subsection{Aggregation Strategies}

We consider three ensemble strategies in this paper: \textit{AveFL}, \textit{OptFL} and \textit{MixFL}. As we discussed in the previous section, \textit{AveFL} averages the parameters of models during ensemble; \textit{OptFL} opts for the model with optimal performance; and \textit{MixFL} is the mixture of \textit{AveFL} and \textit{OptFL}, in each iteration choose the better one between them. 
The result, shown in Figure~\ref{fig4}, demonstrate that there is no big difference between \textit{OptFL} and \textit{MixFL} for both cases of 10 and 20 acquisitions. However, the whole distribution they learned is different, as we discuss in Section~\ref{cen_decen}.


\subsection{Non-IID Data Type~\rom{2} (FL with AL)}

\begin{figure}
\centerline{\includegraphics[width=60mm]{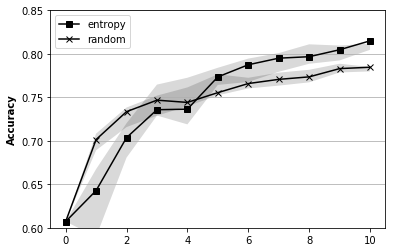}}
\caption{Given the same number of data, AL outperforms random choice in terms of accuracy.}
\label{al}
\end{figure}

\begin{figure}[t!] 
\begin{subfigure}{0.249\textwidth}
\includegraphics[width=\linewidth]{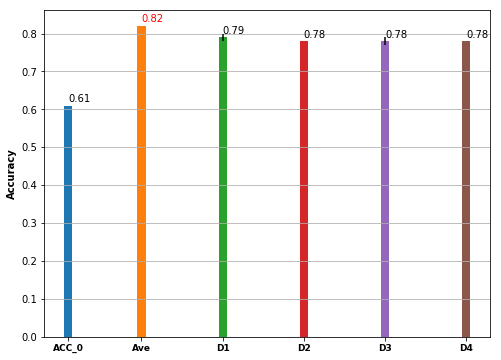}
\caption{Random} \label{al_fl_fre:a}
\end{subfigure}\hspace*{\fill}
\begin{subfigure}{0.249\textwidth}
\includegraphics[width=\linewidth]{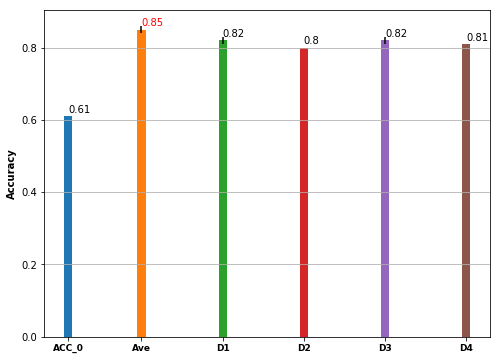}
\caption{Entropy} \label{al_fl_fre:b}
\end{subfigure}
\caption{From left to right we plot initial accuracy, ensemble accuracy and four model accuracy without ensemble step. First of all, no matter random or AL, the result of ensemble model has higher accuracy compared with no ensemble. Overall, AL has better performance with respect to random choice (from the second bar to the last bar).}
\label{al_fl}
\end{figure}

For Non-IID Type \RN{2}, we simulate it by applying AL on the four edge devices. AL can be considered an effective way of choosing data than random sampling, and this behavior causes a slightly biased data generation. For instance, in \cite{qian2019active}, we select the data with maximal entropy (uncertainty) to train our model. The method is shown in Algorithm \ref{non-iid2}. In Figure~\ref{al}, we firstly show that AL outperforms random choice in terms of prediction accuracy. Also, in Figure~\ref{al_fl} shows how aggregation affects the performance when FL combines AL. . 

\begin{figure*}[t!] 
\begin{subfigure}{0.45\textwidth}
\includegraphics[width=\linewidth]{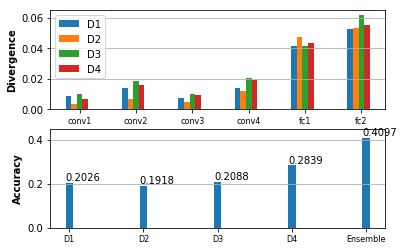}
\caption{Divergence Level \rom{1}} \label{fig:a}
\end{subfigure}\hspace*{\fill}
\begin{subfigure}{0.45\textwidth}
\includegraphics[width=\linewidth]{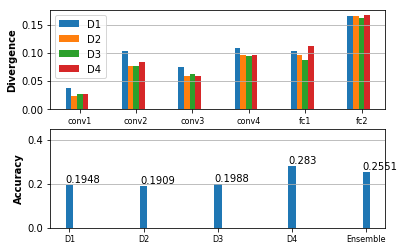}
\caption{Divergence Level \rom{2}} \label{fig:b}
\end{subfigure}
\begin{subfigure}{0.45\textwidth}
\includegraphics[width=\linewidth]{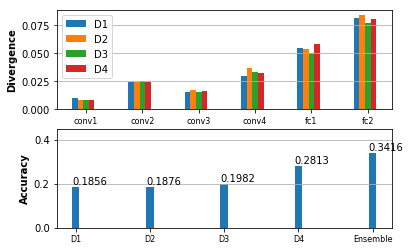}
\caption{Divergence Level \rom{3}} \label{fig:c}
\end{subfigure}\hspace*{\fill}
\begin{subfigure}{0.45\textwidth}
\includegraphics[width=\linewidth]{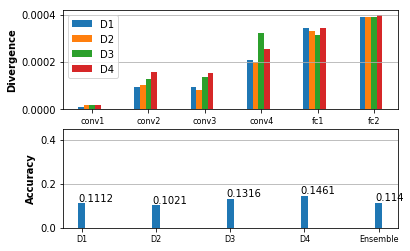}
\caption{Divergence Level \rom{4}} \label{fig:d}
\end{subfigure}
\caption{Plot above represent the divergence and plot below is the corresponding accuracy of local models and ensemble model. X coordinator of plot (above) indicates the layers of model and bars with different colors represent the different devices. Ensemble Effect has high correlation with divergence grade. Here we compared four level of divergence, From Fig \ref{fig:a} to Figure \ref{fig:d} the divergence decreases, Figure \ref{fig:c} corresponds to best ensemble performance. Note that y coordinators are not aligned due to the different magnitudes.}
\label{divergence}
\end{figure*}

\begin{figure}[t!] 
\begin{subfigure}{0.25\textwidth}
\includegraphics[width=\linewidth]{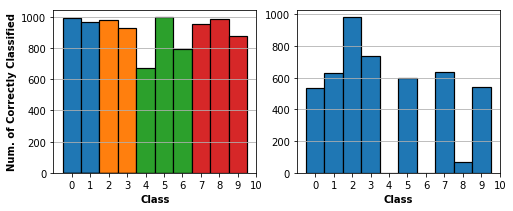}
\caption{\textit{AveFL}} \label{hist:a}
\end{subfigure}\hspace*{\fill}
\begin{subfigure}{0.25\textwidth}
\includegraphics[width=\linewidth]{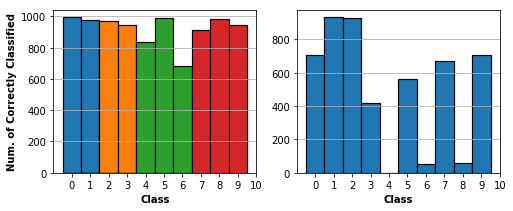}
\caption{\textit{MixFL}} 
\label{hist:b}
\end{subfigure}
\caption{Figure \ref{hist:a} is the correctly classified histogram of Average ensemble case, in the left plot the four different color bars correspond to four models of edge devices without ensemble. They can only correctly classify their own categories. While the right plot in Figure \ref{hist:a} has a better comprehensive capability of classification, it has the difficulty of distinguishing classes 4 and 6. In Figure \ref{hist:b} we plot the same results for the mix ensemble method.}
\label{hist}
\end{figure}

\begin{figure}[t!] 
\begin{subfigure}{0.25\textwidth}
\includegraphics[width=\linewidth]{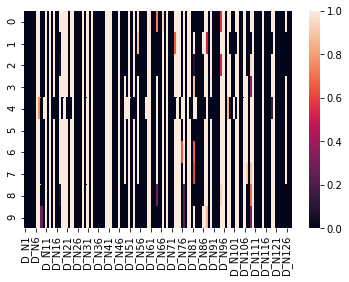}
\caption{Activation of Model from D1} \label{heat:a}
\end{subfigure}\hspace*{\fill}
\begin{subfigure}{0.25\textwidth}
\includegraphics[width=\linewidth]{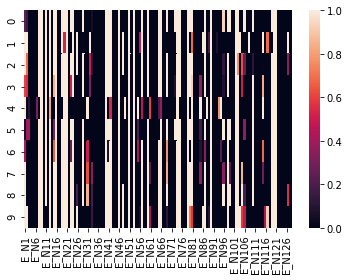}
\caption{Activation of Ensem. Model} \label{heat:b}
\end{subfigure}
\centering
\begin{subfigure}{0.25\textwidth}
\includegraphics[width=\linewidth]{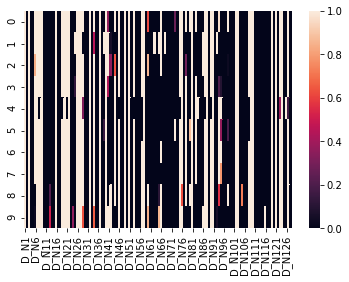}
\caption{Activation of Model from D3} 
\label{heat:c}
\end{subfigure}
\caption{We choose 10 test images from class one and feed into local model from D1 in Figure \ref{heat:a}, ensemble model in Figure \ref{heat:b} and local model from D3 in \ref{heat:c}. The x coordinator indicates the nodes of the fully connected layer right before the softmax layer, and the heatmap values are the output of the nodes. The activation of pattern of D1 is similar to the ensemble model, fairly different from D3 where class 1 is not generated.}
\label{heatmap}
\end{figure}

\subsection{Weights Divergence and Aggregation Performance}\label{weight_divergence}
In this subsection, we quantitatively investigate the correlation between weight divergence and ensemble performance. Firstly, let us define the divergence of layer $l$ of device $i$ as follows:
\[
\text{divergence}_i^l=\frac{|W_{i}^{tl} - W_{\text{ensemble}}^{tl}|}{|W_{i}^{tl}|}
\]
Where $W_{i}^{tl}$ is the layer $l$ of model (device) $i$ at iteration $t$ and $W_{\text{ensemble}}^{tl}$ is the the layer $l$ of ensemble model at iteration $t$.\\
In Figure \ref{divergence}, we plot the divergence of all layers in the network (above) and the corresponding aggregated result (below). The first coordinate represents the network layers, and the four colored bars represent the different devices. From Figure \ref{fig:a} to Figure \ref{fig:d}, the divergence decreases for all the layers; however, the aggregation increases in the beginning and stops increasing at some point. It could indicate that if the divergence value is too large (Figure \ref{fig:a}), the aggregation effect is not fully enabled, and on the other hand, if it is too small (Figure \ref{fig:d}), it may disable the aggregation effect. Our result is consistent with \cite{zhao2018federated}, where they also showed the accuracy reduction is significantly correlated with weight divergence.

\subsection{Correlation between the device models and Aggregated model}\label{cen_decen}
We can also consider the ensemble model as the Gaussian Mixture Model (GMM) \cite{verbeek2003efficient}. Suppose we have $C$ classes, which correspond to $C$ models $M_1,M_2,..,M_C$. We define GMM as $M_{GMM} = \sum_{i=1}^C \alpha_i M_i$ and $\sum_{i=1}^C \alpha_i = 1$. In our case, we consider $\alpha$ uniformly distributed since we do not have prior knowledge and do not want to solve the optimization problem to compute $\alpha_i$. It implies an assumption that the ten classes share some common features, otherwise, it does not make sense. Averaging weights is like partially `copying' the classification capability of different classes from their related edge devices. We call it `partially' because the effect will be diluted by models from other categories. The experimental evidence is shown in Figure \ref{hist}. For \textit{AveFL} (Figure \ref{hist:a}), we plot the histogram of correctly classified classes for four edge devices (corresponds to four colors) in the left figure and the histogram of correctly classified classes for the average model in the right figure. As we can see, without ensemble the local model can \textbf{only} predict the categories generated from their own. For instance, $D_1$ generates class $0$ and $1$, the trained model on D1 can only predict \textbf{$0$} and \textbf{$1$}. However, the ensemble model can predict most of the classes, except with the difficulty to classify $4$ and $6$. In Figure~\ref{fmnist}, we can see class $4$ is `coat' and class $6$ corresponds to `shirt'(label starts from $0$). These two classes are very similar, and it is difficult to distinguish, particularly when the minor different between two classes has been neutralized after ensemble. While, \textit{MixFL} (Figure~\ref{hist:b}) has different behavior: it learns different distributions from ensemble, though, their overall accuracy is similar (shown in Figure~\ref{fig4}).\\ 
To further study how  local models benefit the ensemble model, we analyze the neuron activation patterns. We choose 10 test images from class 1 and feed them into local model from D1 in Figure~\ref{heat:a}, ensemble model in Figure~\ref{heat:b} and local model from D3 in Figure~\ref{heat:c}. The x coordinator indicates the nodes of the fully connected layer right before the softmax layer, and the heatmap values are the output of the nodes. The activation of pattern of D1 is similar to the ensemble model, fairly different from D3 that does not have any clue about class 1. 

\section{Conclusion}\label{sec:Conc}
Distributed machine learning has several virtues, including the potential to reduce data aggregation and thus improved privacy. However, this virtue poses a potential challenge, namely that the edge devices are set to learn from Non-IID data. Hence, to investigate the robustness of Federated Learning to Non-IID data, we simulate two scenarios. Furthermore we analyze and compare different aggregation strategies: \textit{AveFL}, \textit{OptFL} and \textit{MixFL}. We presented evidence that federated learning is robust to sampling bias, and also we found that the epoch (amount of local learning) and the aggregation frequency are important parameters for Federated Learning. In the end, we also post-process the prediction performance to understand the correlation between local models and the aggregated model.\\

\clearpage
\bibliographystyle{IEEEtran}
\bibliography{references}

\end{document}